\documentclass{article}
\usepackage[utf8]{inputenc}
\usepackage[margin=1in]{geometry}
\usepackage{graphicx} % Required for inserting images
\usepackage{subcaption}  % for modern subfigure support

\usepackage{graphicx}%
\usepackage{multirow}%
\usepackage{amsmath,amssymb,amsfonts}%
\usepackage{amsthm}%
\usepackage{mathrsfs}%
\usepackage[title]{appendix}%
\usepackage{xcolor}%
\usepackage{textcomp}%
\usepackage{manyfoot}%
\usepackage{booktabs}%
\usepackage{algorithm}%
\usepackage{algorithmicx}%
\usepackage{algpseudocode}%
\usepackage{listings}%
\usepackage{xcolor}
\usepackage{soul}
\usepackage{makecell}
\usepackage{gensymb}
\usepackage{mathtools}
\usepackage{setspace}
\usepackage{authblk}
\usepackage{wasysym}
\usepackage{physics}
\usepackage{url}
\usepackage{natbib}

\newcommand{\bl}[1]{\boldsymbol{#1}}

\title{Turbulent oscillation in unbalanced T-junction flows}

\author[1]{Dongjie Jia\thanks{Email: jia191@purdue.edu}}
\author[1]{Arezoo Ardekani\thanks{Correspondence: ardekani@purdue.edu}}
\affil[1]{School of Mechanical Engineering, Purdue University, West Lafayette, Indiana, USA}

\begin{document}
\maketitle

\begin{abstract}
The T-junction impinging flow occurs in many fluid dynamics systems.
In particular, the T-junction micromixer has recently been widely used for nanoparticle production, where the two inlet streams operate at a significant flow-rate imbalance and the Reynolds number is in the turbulent regime. 
This operating condition exposes a gap in the existing literature on the fluid dynamics of the T-junction.
In this study, we used high-fidelity numerical simulations to investigate high-Reynolds-number unbalanced T-junction flows.
We discover a new oscillatory behavior between the two inlet streams at the T-junction, leading to a new turbulence-production mode. 
We will present detailed evidence of this new behavior, in contrast to the existing understanding of balanced turbulent T-junction flows.
This oscillatory behavior also persists across a range of Reynolds numbers simulated, where the Strouhal number is approximately constant, indicating a self-similar phenomenon. 
As a result, many of the fluid dynamics parameters follow a power-law relation with the Reynold number.
The discovery in this paper affects real-world applications, where process design and product quality are affected by turbulence and mixing dynamics. 

\end{abstract}

%{\bf MSC Codes }  {\it(Optional)} Please enter your MSC Codes here

\section{Introduction}
Two flows impinging in a T-shaped junction is a common occurrence in many systems, such as heat exchangers, ventilation systems, and microfluidic mixers, with a wide range of applications including nuclear power plants, chemical processing, and nanoparticle synthesis.
Understanding the fluid dynamics of the T-junction has always been essential for the proper design and application of these devices.
Existing research classifies the fluid dynamics behavior in the T-junction into four regimes in the order of increasing Reynolds number ($Re$): segregated, vortex, and engulfment flows, which are time-steady \citep{engler2004numerical,soleymani2008numerical,orsi2013water}, and the unsteady turbulent flow \citep{schikarski2017direct}. 
The bulk of existing numerical and experimental literature focuses on the steady flow regimes where the Reynolds number is typically less than 400.  

In recent years, the T-junction micromixer gained increasing interest for its application in nanoparticle production, such as COVID-19 mRNA vaccines \citep{chen2025controlled}. 
This application is based on the principle of nanoprecipitation \citep{rivas2017nanoprecipitation}, where an organic solvent (ethanol) carrying hydrophobic lipids rapidly mixes with a non-solvent, typically water, to form lipid nanoparticles. 
This process requires the T-junction mixer to operate at a flow condition beyond the existing knowledge base. 
To begin with, the micromixer operates in the turbulence regime ($Re\sim O(1000)$) to ensure a mixing time scale smaller than the nanoparticle formation time scale. 
In addition, the antisolvent-to-solvent flow rate ratio is typically 3:1 to prevent nanoparticle aggregation, resulting in significant flow imbalance.
Furthermore, the fluid properties of the two streams are different, adding additional complexity. 

In this study, we use the stabilized finite element methods to investigate the unbalanced T-junction flow with water-ethanol mixtures at intermediate to high Reynolds numbers in the turbulence regime. 
The turbulent flow is resolved using the residual-based variational multiscale method, which is analogous to high-fidelity large eddy simulation and has analytically derived error bounds \citep{bazilevs2007variational}.  
The mesh element length scale used in the simulations is on the same order of magnitude as the Kolmogorov length scale, ensuring close to direct numerical simulation accuracy while keeping the simulation cost reasonable. 
To our knowledge, this is the first study to investigate turbulent unbalanced T-junction flow, in which we discovered a new unsteady interfacial oscillation behavior, leading to a new turbulence-production mode.
This discovery is of importance for both fundamental understanding of the T-junction flows and the process and formulation design for nanoparticle production. 

\section{Method}

The physical behavior in this study is governed by the compressible Navier-Stokes equations and the mass-transport equation. The stabilized finite element method is used to solve the governing equations. The mixer geometry is scaled to simulate a range of Reynolds numbers. 
In this section, we will detail the governing equations, the numerical implementation, and the simulation setup.

\subsection{Governing equations}

To solve for the fluid dynamics of two miscible liquids, water and ethanol, we start from the Navier-Stokes equation.  
Assuming isothermal flow with no external forces, the Navier-Stokes equations is given as
\begin{equation}
\left. \begin{array}{ll}
    \rho\frac{\partial\bl u}{\partial t} + \rho\bl u \cdot \bl\nabla{\bl u} + \bl\nabla p - \mu\bl\nabla^2{\bl u} - \bl\nabla\mu \cdot \left[\bl\nabla \bl{u} +(\bl\nabla \bl {u} )^{\intercal}\right] = 0,  \\[8pt]
    \frac{\partial\rho}{\partial t}  + \rho\bl\nabla\cdot  \bl u +\bl\nabla\rho\cdot\bl u = 0,  \\
 \end{array}\right\}
 \label{eqn:NS-og}
\end{equation}
where $\bl u$ is the fluid velocity vector and $p$ is the pressure. $\rho$ and $\mu$ are the density and dynamic viscosity of the fluid, respectively. 
Although both water and ethanol are incompressible, the total fluid volume is not conserved because of hydrogen bonding and molecular packing effects when water and ethanol mix. 
Therefore, the flow field is not divergence-free, i.e., $\nabla\cdot\bl u\neq0$. 

The mixing dynamics is described by the mass transport equation as
\begin{equation}
    \frac{\partial w}{\partial t}  + \bl u\cdot\bl\nabla w -D \bl\nabla^2w-\frac{D}{\rho}\bl\nabla \rho\cdot\bl\nabla w = 0,
    \label{eqn:AD-og}
\end{equation}
where $w$ is the mass fraction of ethanol. $D=1.23\times10^{-5}\:\mathrm{cm^2/s}$ is the diffusivity of ethanol in water. 
Note that the last term is non-negligible due to the non-uniform density. 

\begin{figure} [ht]
  \centerline{\includegraphics{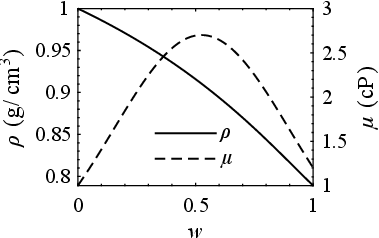}}% Images in 100% size
  \caption{Water-ethanol mixture density, $\rho$, (solid line, left axis) and dynamic viscosity, $\mu$, (dashed line, right axis) as functions of ethanol mass fraction ($w$).}
\label{fig:rhomu}
\end{figure}

As shown in figure \ref{fig:rhomu} in a solid line, the fluid density, $\rho$, is calculated based on a Jouyban-Acree model \citep{jouyban2018mathematical} for water-ethanol mixtures \citep{khattab2012density} at room temperature $T = 293\:\mathrm{K}$ as
\begin{equation}
    \rho = \exp\left[(1-x)\ln(\rho_w)+x\ln(\rho_e)-\frac{30.8}{T}(1-x)x\right],
    \label{eqn:rho}
\end{equation}
where $\rho_w = 1\:\mathrm{g/cm^3}$ and $\rho_e = 0.789\:\mathrm{g/cm^3}$ are the density of water and ethanol, respectively. 
The mole fraction of ethanol, $x$, is calculated from the ethanol mass fraction, $w$, as
\begin{equation}
    x = \frac{M_w w}{M_e(1-w)+M_w w},
    \label{eqn:mass-mole}
\end{equation}
where $M_w = 18.02\:\mathrm{g/mol}$ and $M_e = 46.07 \:\mathrm{g/mol}$ are the molar masses of water and ethanol, respectively. 

The dynamic viscosity, $\mu$, is calculated using a Grunberg-Nissan model \citep{grunberg1949mixture} fitted to experimental data \citep{furukawa2017brownian} as
\begin{equation}
    \mu = \exp\left[(1-w)\ln(\mu_w) + w\ln(\mu_e) + 3.6(1-w)w\right],
    \label{eqn:mu}
\end{equation}
where $\mu_w = 0.01 \:\mathrm{cP}$ and $\mu_e = 0.012 \:\mathrm{cP}$ are the dynamic viscosity of water and ethanol, respectively. 
The dynamic viscosity curve is shown in figure \ref{fig:rhomu} in a dashed line.

\subsection{Numerical implementation}
The above governing equations are numerically solved using an in-house solver named Multiphysics Finite Element Solver (MUPFES) \citep{moghadam2013modular,esmaily2013new}. 
The solver has previously been validated \citep{esmaily2015bi} and deployed in studies on turbulent flows \citep{jia2021simulation,jia2022characterization}.
The exact details of the numerical framework used in this study are available in our previous publication \citep{jia2026high}.

The stabilized finite element method is used to mitigate numerical oscillations and allows the use of equal shape functions in advection-dominant regimes \citep{hughes1986new}, as is the case of this study.
We use the residual-based variational multiscale method (RBVMS) to stabilize the Navier-Stokes equations and model the sub-grid scale turbulence \citep{bazilevs2007variational,bazilevs2008isogeometric}.
This method performs a multi-scale decomposition to the velocity and pressure fields and models the subgrid-scale velocity and pressure based on the residuals of the Navier-Stokes equations.
For the mass transport equation, the streamline upwind Petrov–Galerkin (SUPG) method \citep{brooks1982streamline} is used to stabilize the weak-form. 
Similarly to the method used for the Navier-Stokes equations, this method includes an artificial diffusion term calculated from the numerical residual \citep{jia2023time}.

For time integration, the solver uses a second-order generalized-$\alpha$ method \citep{jansen2000generalized}, where $\rho_{\infty} = 0.2$.
At each time step, the systems of equations are linearized and iteratively solved using a weakly-coupled Newton-Raphson (NR) method.
At each Newton-Raphson iteration, the linear systems for the Navier-Stokes equations and the advection-diffusion equation are constructed from the latest unknown values and solved sequentially.
% The Navier-Stokes equations is solved first with the density and viscosity are calculated from the ethanol mass fraction of the previous NR iteration. 
% The advection-diffusion equation is then constructed and solved using the current NR iteration velocity. 
At each time step, the Newton-Raphson iteration stops after both equations' residuals are reduced by four orders of magnitude. 
This implementation ensures simultaneous convergence for both equations while maintaining reasonable the linear solver cost.

The linear system for the Navier-Stokes equations is solved using a specialized bi-partitioned iterative algorithm \citep{esmaily2015bi}.
The linear system for the advection-diffusion equation is solved using a preconditioned generalized minimal residual (GMRES) algorithm \citep{shakib1989multi}.
The tolerances are set to 0.4 and 0.1 for the Navier-Stokes linear solver and GMRES, respectively.
At each time step, the ethanol mass fraction, $w$, is numerically clipped to its physical bound on $[0,1]$ to prevent numerical divergence. 
We have shown that such an implementation has a negligible effect on the overall mass conversation \citep{jia2026high}.

\subsection{Simulation setup} 

\begin{figure} [ht]
  \centerline{\includegraphics{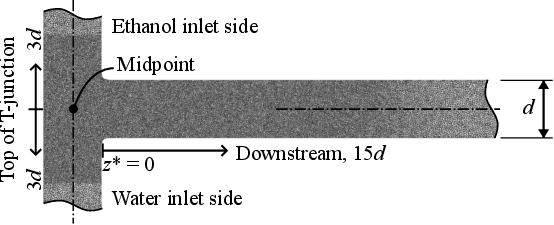}}% Images in 100% size
  \caption{Center plane view of the T-junction geometry and mesh. Portion of the inlets and downstream are clipped (inidicated by $\sim$). The axial of the inlets and downstream are marked in dot-dashed lines. $d$, diameter; $z^*=z/d$ non-dimensional downstream direction. }
\label{fig:mesh}
\end{figure}

The T-junction geometry consists of two collinear cylindrical inlets opposing each other and a cylindrical downstream pipe perpendicular to the inlets at the midpoint of the two inlets. 
The diameter, $d$, of the inlet and outlet pipes is constant throughout the geometry.
The total length of the inlets is six times the diameter, $6d$, resulting in $3d$ for each inlet from the midpoint.
The length of the downstream pipe is fifteen times the diameter, $15d$. 
The edge between the inlets and downstream is filleted by $d/10$ to ensure a smoother flow transition and to prevent numerical instability caused by a sharp edge. 

The mesh is created using an open-source package named \textsc{TetGen} \citep{hang2015tetgen}.
Initially, a $d=0.04$ cm T-junction geometry is meshed using unstructured linear tetrahedral elements with a maximum element volume constraint of $5\times10^{-10}\: \mathrm{cm^3}$, resulting in element edge lengths on the order of $10^{-3} \:\mathrm{cm}$.
The center region of the mesh, where the fluid impingement occurs and turbulence develops, is then further refined. 
More specifically, regions $1.1d$ on each side of the inlet from the midpoint and $5d$ downstream are remeshed with twice the edge resolution, or eight times the volumetric resolution. 
Elements in the region from $5d$ to $10d$ downstream are linearly refined from the fine mesh to the original mesh, to ensure a gradual change in element size. 
The final mesh contains $6,273,321$ elements.

For all cases, the total flow rate is fixed at 80 mL/min. 
For the balanced flow rate case, both the water and ethanol inlets run at 40 mL/min.
In the unbalanced flow rate case, the water stream runs at 60 mL/min, while the ethanol stream runs at 20 mL/min, resulting in a water-to-ethanol flow rate ratio (FRR) of 3:1.
This flow rate ratio is in line with the condition used for the production of pharmaceutical drugs. 
For the Navier-Stokes equations, the wall is no-slip, and the outlet is a zero-Neumann boundary condition.
Due to the numerical back-flow stabilization used at the outlet, solutions within one diameter of the outlet are omitted from the results to ensure physical accuracy. 
For the mass transport equation, $w=0$ is imposed at the water inlet, and $w=1$ is imposed at the ethanol inlet. 
The walls and outlet are zero Neumann boundary conditions. 
The Reynolds number is varied by changing the diameter while keeping all other parameters fixed.
The mesh geometry scales uniformly with the change in diameter.

All simulations are run in parallel using 125 cores at 2.2GHz on Purdue University's Negishi cluster with AMD Epyc ``Milan" processors. 
Each simulation takes around 30 hours to reach statistical convergence.
The solutions are then saved for a minimum of five oscillation periods. 

\section{Results}

In this section, we will describe the newly discovered oscillation behavior and turbulence production mode for the unbalanced T-junction flow. We will characterize the turbulence and mixing dynamics across a range of Reynolds numbers and discuss the implications for nanoparticle production. 
The definition and scalings of the non-dimensional numbers and variables are given in appendix \ref{appA}. 
In the results, $\overline{\cdot}$ indicates the spatial average, on a cross-sectional surface, and $\langle\cdot\rangle$ indicates the temporal average.
All units are in centimeter–gram–second (CGS).

\subsection{Turbulence and oscillation}

\begin{figure} [ht]
  \centerline{\includegraphics{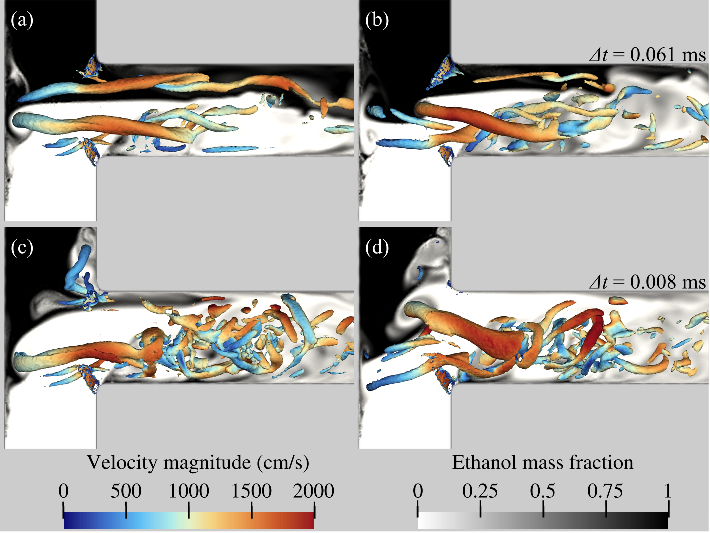}}% Images in 100% size
  \caption{Q-criterion isosurfaces ($Q = 2\times10^{10}$) colored by the velocity magnitude, plotted over the ethanol mass fraction at the center plane. (a) and (b) are two snapshots from the balanced flow rate result ($Re=3451$). (c) and (d) are two snapshots from the unbalanced flow rate result ($Re=3829$). $\Delta t$ is the time gap between the left and right snapshots.}
\label{fig:11v13}
\end{figure}

To contrast the turbulence production and evolution differences between balanced ($Re=3451$) and unbalanced flows ($Re=3829$), we first show the results of the two cases running at the same total flow rate ($Q = 80\:\mathrm{mL/min}$) in figure \ref{fig:11v13}. 
At balanced inlet flow rates, similar to observations in existing experimental and numerical studies \citep{schikarski2017direct,zhang2019investigation}, the turbulence is generated through two phenomena: 1. periodic formation and shedding of a vortex pair at the top of the T-junction and 2. destabilization and breakage of the interfacial plane in the downstream.
As shown in figure \ref{fig:11v13}(a), a fully formed vortex pair originates from the recirculation zone at the top of the T-junction.
In figure \ref{fig:11v13}(b), this vortex pair detaches and propagates downstream in the axial direction, while at the same time a new vortex pair begins to form at its previous location. 
It is important to note that although the flow rate is balanced, the densities and viscosities of water and ethanol are different.
Therefore, the turbulent structure is not fully symmetric, with slightly stronger turbulence on the water side. 

On the other hand, when the flow rate is unbalanced, with a 3-to-1 flow rate of water to ethanol,  a single vortex develops at the top of the T-junction on the water side of the inlet channel, as shown in figure \ref{fig:11v13}(c).
This vortex tube then detaches from the wall and propagates along the local flow direction, which contains both a downstream component and a tangential component towards the ethanol side, shown in figure \ref{fig:11v13}(d).
This tangential velocity component of the detached vortex causes it to collide with the wall on the ethanol side. 
This collision violently splatters the vortex tube, generating intense turbulence that rapidly dissipates energy. 
The remanent structures of the collision can be seen in figure \ref{fig:11v13}(c), both on the ethanol side of the inlet and in the downstream.
The single vortex formation and detachment rate is significantly faster than that of the vortex pair under balanced flow conditions. 

The single vortex shedding creates an oscillating interface between the two fluids under unbalanced flow. 
As the vortex strengthens and detaches from the wall, the trailing edge of the vortex pulls the ethanol streams towards the water side, as shown in figure \ref{fig:11v13}(d).
Once the vortex tube collides with the wall and splatters, the higher momentum of the water stream pushes the interface back towards the ethanol side as the next vortex tube forms, as shown in figure \ref{fig:11v13}(c). 
These effects create an interface oscillation in the inlet axial direction, leading to much earlier breakup of the interfacial plane in the unbalanced flow than in the balanced flow. 

\begin{figure} 
  \centerline{\includegraphics{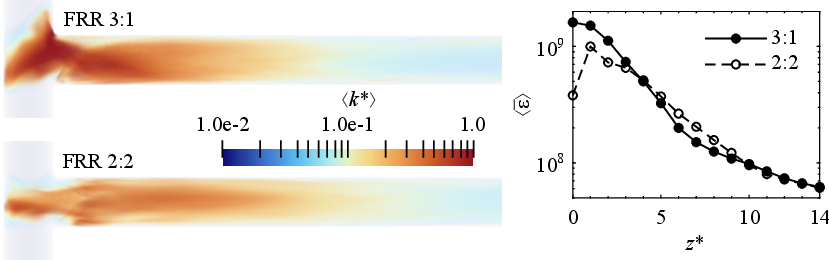}}% Images in 100% size
  \caption{Left: time-averaged non-dimensional turbulent kinetic energy, $\langle k^*\rangle$, volumetric contour colored in log-scale. Right: mean turbulent dissipation rate, $\overline{\varepsilon}$, along the non-dimensional downstream $z^*$, with a y-axis range of $[5\times10^7, 2\times10^9]$. Unbalanced flow (FRR 3:1, solid line, $Re=3829$); balanced flow (FRR 2:2, dashed line, $Re=3451$)}
\label{fig:tked4}
\end{figure}

The newly discovered single-vortex collision behavior in the unbalanced flow significantly affects turbulence production and dissipation, as shown in figure \ref{fig:tked4}.
The non-dimensional turbulent kinetic energy, $k^*$, is scaled by the characteristic inlet kinetic energy, defined in ~\ref{eqn:k}.
For the unbalanced flow, the turbulence kinetic energy production is instantaneous, peaking at the T-junction, whereas for the balanced flow, the turbulence undergoes a development period and the energy peaks at one to two diameters downstream. 
Moreover, the peak non-dimensional energy in the unbalanced flow is much higher than that of the balanced flow, indicating more efficient turbulence production. 
As shown in figure \ref{fig:tked4}, the difference in turbulence dynamics also affects the turbulence dissipation rate, which is directly computed from the Reynolds stress tensor. 
The differences in both turbulent energy and microscale have important implications for nanoparticle synthesis, which will be discussed in section \ref{sec:implication}.

\subsection{Characterization across Reynolds numbers}

\begin{table}
  \begin{center}
  \begin{tabular}{rrcclllcc}
      $Re$  & $Q$   &   $d$ & $St$   & $\langle\overline{k}\rangle_{z^*=0}$ & $\langle\overline\varepsilon\rangle_{z^*=0}$& $\langle\overline\eta\rangle_{z^*=0}$ & $M_{z^*=14}$ & $z_{M=0.9}^*$ \\[3pt]
       ~479   & ~80 & 0.32  & 0.4766 & $11.39~~~~~~~~~~$   & $1.715\times10^3$ & $1.258 \times 10^{-2}$ & 0.8344 & \--  \\
       ~957   & ~80 & 0.16  & 0.4539 & $3.342 \times 10^2$ & $2.076\times10^5$ & $3.415 \times 10^{-3}$ & 0.9255 & 11.4 \\
       1914   & ~80 & 0.08  & 0.4255 & $6.887 \times 10^3$ & $1.833\times 10^7$ & $1.095 \times 10^{-3}$ & 0.9669 & 7.0  \\
       3829   & ~80 & 0.04  & 0.4231 & $1.355 \times 10^5$ & $1.610\times 10^9$ & $3.526 \times 10^{-4}$ & 0.9729 & 5.4  \\
       7658   & ~80 & 0.02  & 0.4047 & $2.216 \times 10^6$ & $1.186\times 10^{11}$ & $1.163 \times 10^{-4}$ & 0.9759 & 3.7  \\
   $^+ 3829$  & 160 & 0.08  & 0.4103 & $3.413\times 10^4$ & $1.047\times10^8$ & $6.924\times 10^{-4}$ & 0.9776 & 5.3\\
  \end{tabular}
  \caption{Characterization summary for the T-junction operating at 3:1 unbalanced flow. $Q$, total flow rate; $k$, turbulent kinetic energy; $\varepsilon$, turbulent dissipation rate; $\eta$, Kolmogorov length scale; $M$, mixing index. $^+$The additional simulation for validation purposes. } 
  \label{tab:st}
  \end{center}
\end{table}

Under unbalanced flow conditions, five different Reynolds numbers of equal spacing in the logarithmic scale are simulated, and the results are summarized in table \ref{tab:st}.
Since we only changed the length scale to adjust the Reynolds number, an additional simulation was performed at $Re = 3829$ with a different total flow rate and length scale to provide additional validation. 
The relatively constant Strouhal number indicates that the oscillatory behavior observed in this study is self-similar. 
Calculating from the original five cases with the same total flow rate, the mean Strouhal number is $\overline{St} = 0.4367$ and the 99\% confidence interval is $St_{0.99} = [0.4041,0.4694]$.
The onset of the oscillation phenomenon is between $Re = 239$ (result not shown) and $Re=479$. 

\begin{figure}
  \centerline{\includegraphics{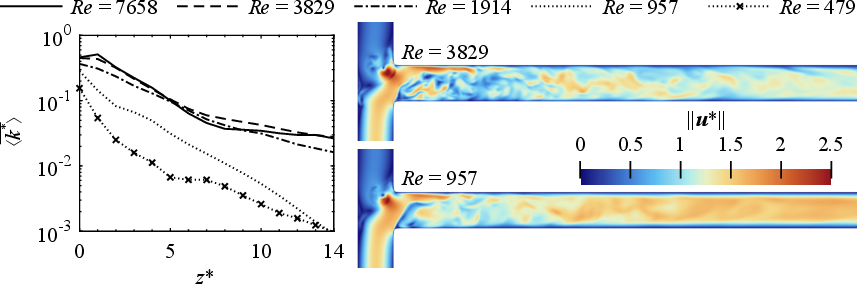}}% Images in 100% size
  \caption{Left: mean non-dimensional turbulent kinetic energy, $\langle\overline{k^*}\rangle$, over the non-dimensional downstream, $z^*$, for all Reynolds numbers. Right: non-dimensional velocity magnitude contour at the center plane.}
\label{fig:tkere}
\end{figure}

Although the oscillation frequency caused by periodic vortex detachment is self-similar across the range of reported Reynolds numbers, the downstream turbulence behavior is more nuanced.
As shown in figure \ref{fig:tkere}, the cross-section mean non-dimensional turbulent kinetic energy is at a similar value at the start of the downstream, $z^*=0$, which is as expected due to the self-similar oscillation mode.
However, the turbulent kinetic energy profile in the downstream is not self-similar until $Re = 1914$.
This indicates that although the oscillatory mode at the junction induces instabilities at a lower Reynolds number $\sim O(100)$, the Reynolds number must reach $\sim O(10^3)$ to sustain this instability downstream. 
As shown on the right in figure ~\ref{fig:tkere}, the $Re=3829$ case exhibits a typical turbulent pipe flow pattern in the downstream, while the low Reynolds number case reverts back to a laminar flow.
This reversion to laminar flow is exacerbated by the non-monotonic behavior of the mixture viscosity (figure \ref{fig:rhomu}), leading to a reduction in $Re$ by approximately half further downstream, when the fluid becomes homogeneous. 

Under this self-similar behavior at the T-junction, as summarized in table \ref{tab:st}, the turbulent kinetic energy, $k$, the turbulent dissipation rate, $\varepsilon$, the Kolmogorov length scale, $\eta$, and mixing distance/time, $z^*_{M=0.9}$, all show power law relations with the Reynolds number.
This predictable scaling is a key advantage of the T-junction in nanoparticle production applications. 

\subsection{Implications for application}
\label{sec:implication}

\begin{figure}
  \centerline{\includegraphics{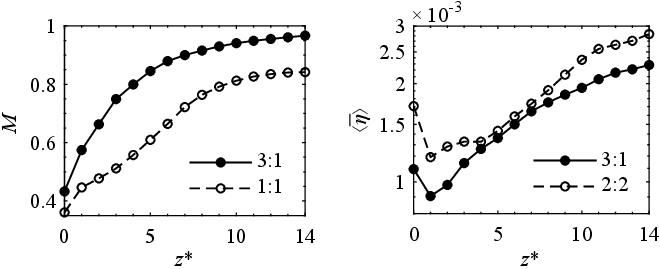}}% Images in 100% size
  \caption{Left: mixing index, $M$, along the downstream direction, $z^*$. Right: mean Kolmogorov length scale, $\langle\overline{\eta}\rangle$, along the downstream direction, $z^*$, with a y-axis range of $[8\times10^{-4}, 3\times10^{-3}]$. Unbalanced flow (3:1, solid line, $Re=1914$); balanced flow (2:2, dashed line, $Re=1725$).}
\label{fig:MI}
\end{figure}

The T-junction mixer is extensively used in pharmaceutical vaccine production. 
In this application, the degree of mixing and mixing scales directly affect the design of the production process and formulation. 
The additional turbulent mode discovered in the unbalanced flow both shortens the mixing time and improves the mixing homogeneity, as quantified by the mixing index $M$ in figure \ref{fig:MI}. 
$M=0$ indicates two immiscible fluids, and $M=1$ indicates a homogeneous mixture.
Due to the more intense turbulence, the mean Kolmogorov length scale is about twice as small for the unbalanced flow as for the balanced flow.
Note that the minimum Komolgorov length scale will be around two orders of magnitude smaller that the mean Komolgrov scale plotted in figure \ref{fig:MI}.
These newly discovered differences in mixing length and time scale need to be accounted for when investigating nanoparticle differences from T-mixers.
Due to the self-similar unbalanced flow and the predictable turbulence and mixing characteristics, it is possible to design experiments to isolate and quantify the effects of each turbulence and mixing parameter on nanoparticle product quality and characteristics. 
With knowledge of potential kinetics-kinematics relations, the T-junction enables straightforward design optimization for specific types of nanoparticles. 

\section{Conclusion}

We present a new turbulence production mode discovered when operating a T-junction geometry at an unbalanced flow rate at intermediate to high Reynolds numbers. 
The new turbulence mode arises from the oscillatory behavior of the two streams' interface at the junction, which occurs at $Re \ge O(100)$.
The Strouhal number of this oscillation is approximately constant across a range of Reynolds numbers, indicating that this phenomenon is self-similar. 
In combination with the classical turbulent pipe flow theory, the fluid dynamics in the whole geometry, including the junction and downstream, becomes self-similar around $Re\sim O(1000)$. 
These behaviors are not observed in balanced T-junction flows or in low Reynolds number unbalanced flows.
The new discovery affects nanoparticle applications, where process design and product quality are directly influenced by turbulence and mixing dynamics.
%%%%%%%%%%%%%%%%%%%%%%%%%%%%%%%%%%%%%%%%%%%%%%%%%

Declaration of Interests: The authors report no conflict of interest.
%\appendix
\begin{appendices}

\section{Scaling and non-dimensional definitions}
\label{appA}
For the fluid system in this study, the characteristic length scale is the diameter, $d$.
The characteristic velocity is calculated from the inlets as $U = (\rho_wU_w+\rho_eU_e)/\rho_c$, where $U_w$ and $U_e$ are the mean velocity at the water and ethanol inlets, respectively. We use the velocity-weighted density and viscosity, $\rho_c=(\rho_wU_w+\rho_eU_e)/(U_w+U_e)$ and $\mu_c=(\mu_wU_w+\mu_eU_e)/(U_w+U_e)$, as the characteristic mixture density and dynamic viscosity, respectively. These values are more representative than the homogeneous mixture density and viscosity since the two fluids are not mixed at the turbulence onset. 
The Reynolds number, $Re$, and Strouhal number, $St$, are defined as
\begin{equation}
    Re = \frac{\rho_cUd}{\mu_c}, \quad St = \frac{fd}{U},
    \label{eqn:re}
\end{equation}
where $f$ is the oscillation frequency of the two fluids interface.
Three dimensionless variables are used in this study.
The non-dimensional downstream direction, $z^*$, the non-dimensional velocity, $\bl u^*$, and the non-dimensional turbulent kinetic energy, $k^*$, which are defined as
\begin{equation}
   z^* = \frac{z}{d}, \quad
  \bl u^* = \frac{\bl u}{U}, \quad
  k^* =\frac{k}{K}= \frac{\left<(u'')^2\right>+\left<(v'')^2\right>+\left<(w'')^2\right>}{(\rho_wU_w^2+\rho_eU_e^2)/\rho}.
  \label{eqn:k}
\end{equation}
$\left<\cdot\right>$ indicates the temporal average. Note that since the density field in this study is a variable, Favre averaging is used to compute the turbulence fluctuation velocity, $\bf u''$ \citep{Favre1992}.

\end{appendices} %\clearpage

\bibliographystyle{jfm}
\bibliography{jfm}

\end{document}